\documentclass[11pt]{article}

\usepackage[final]{acl}

\usepackage{times}
\usepackage{latexsym}

\usepackage[T1]{fontenc}

\usepackage[utf8]{inputenc}

\usepackage{microtype}

\usepackage{inconsolata}

\usepackage{graphicx}

\usepackage{booktabs}
\usepackage{multirow}
\usepackage{graphicx}
\usepackage{subcaption}
\usepackage{amsmath}

\title{Is Prosody Lost in Translation?\\ Fine-Grained Cross-Lingual Prosody Similarity Across Languages}

\author{
  Haopeng Xie,
  Ismail Rasim Ulgen,
  Sofia Son,
  Berrak Sisman,
  Philipp Koehn \\
  Center for Language and Speech Processing (CLSP), Johns Hopkins University \\
  Baltimore, MD, USA \\
  \texttt{\{hxie25,sson16\}@jh.edu} \\
  \texttt{\{iulgen1,sisman,phi\}@jhu.edu}
}

\begin{document}
\maketitle
\begin{abstract}
Prosody plays an important role in speech translation, conveying information such as emphasis, emotion, and intent beyond lexical content. However, despite recent progress in expressive speech-to-speech translation (S2ST), little is known about how prosodic patterns are similar/different across languages. Understanding these cross-lingual similarities and differences is crucial for effectively incorporating prosody into expressive S2ST systems. In this work, we present the first fine-grained cross-lingual analysis of prosody using multilingual dubbing data across English-German, English-Spanish, and English-French language pairs. We analyze the similarity of pitch, energy, and temporal feature patterns between source and target speech and investigate the linguistic and alignment-related factors affecting this similarity. Our analysis reveals inherent cross-lingual correlations in prosodic structure between certain languages. The findings provide important insights into the transferability of prosody across languages and offer empirical guidance for future expressive speech-to-speech translation systems.
\end{abstract}

\section{Introduction}

Prosody conveys important information beyond lexical meaning, including emphasis, emotion, intent, and discourse structure \cite{prosody_importance-10.1093/oxfordhb/9780198791768.013.30, prosody_importance-article}. In particular, pitch defines intonation patterns that can substantially affect interpretation, while energy often contributes to emphasis, prominence, and emotional expression. For example, expressions such as ``oh my god'' in English or ``dios mío'' in Spanish may convey surprise, disappointment, frustration, or disbelief depending on pitch and energy patterns rather than lexical content alone. Translations that preserve lexical meaning but fail to preserve these prosodic cues may therefore misrepresent the intended meaning, emotional expression, or emphasis. As a result, effective speech-to-speech translation (S2ST) systems must model not only lexical content but also the prosodic characteristics of speech \cite{prosody_importance-Gussenhoven_2004}.

Prosody is often ignored or heavily simplified in existing S2ST systems~\cite{french_ssml-ouali2025improving, prosody_traditional_s2st_discard_prosody-7792638, s2st_direct-jia2022translatotron, s2st_direct-jia2019direct, mos_analysis-jia2018transfer, prosody_ignore-inaguma2023unity, prosody_ignore-kim2024textless, prosody_ignore-lee-etal-2022-direct}. Most approaches primarily focus on translating lexical content, while the prosodic characteristics of the source speech are either discarded or only indirectly incorporated. One major reason for this limitation is the scarcity of prosodically parallel multi-lingual speech data, where utterances are aligned not only in terms of lexical content but also prosody. As a result, the role and transferability of prosody in speech-to-speech translation is largely underexplored.

Recent expressive S2ST and prosody transfer systems have attempted to transfer or control source-language prosody in translated speech \cite{prosody_transfer_study-swiatkowski2023cross, rasim-passt}. These approaches implicitly assume that at least some prosodic structure exhibits cross-lingual correspondence, enabling expressive characteristics of the source speech to be preserved after translation. However, the extent to which prosodic patterns are actually correlated across languages remains poorly understood. Better understanding of these cross-lingual prosodic correlations may provide valuable insights for designing S2ST systems that more effectively transfer source prosody into translated speech.

Existing studies \cite{brannon2023dubbing, avila2023dialog, prosody_korean_study-zhou2024prosody} investigating prosodic relationships between languages are relatively few and are typically limited to coarse-grained analysis at the utterance level. In practice, however, many prosodic cues that affect meaning and expressiveness occur at much finer temporal resolutions. For this reason, it is crucial to understand fine-grained cross-lingual prosodic correlations in order to model the prosodic cues that can substantially affect the meaning, emphasis, and emotional expression of spoken utterances and their translations.

In this work, we perform a fine-grained cross-lingual analysis of prosody using multi-lingual dubbing data. We focus on empirically characterizing
cross-lingual prosodic correspondence. By combining established speech
processing techniques with professionally dubbed multilingual speech,
we aim to answer fundamental questions about whether, where, and to
what extent prosodic information is preserved across translation. Professional dubbing aims to preserve not only the semantic content of the original speech but also its expressive and emotional characteristics while adapting to the linguistic constraints of the target language. This makes dubbing data a particularly valuable resource for studying prosody transfer between languages in realistic expressive speech settings.

We present the first large-scale fine-grained analysis of prosody across English-German, English-Spanish, and English-French language pairs using professionally dubbed speech. Unlike prior work that primarily focuses on utterance-level statistics or monolingual analysis, we investigate how local expressive patterns are preserved during translation at a much finer temporal resolution. Specifically, we analyze cross-lingual similarities in fine-grained pitch and energy using alignment-aware word-level comparison. Additionally, we analyze temporal prosody patterns between different languages to provide further insights for prosody correlation.

Beyond measuring overall prosody correlation, we analyze the effect of word lexical subgroup on prosody transfer. Our analysis provides valuable insights regarding aspects of prosody that are similar/dissimilar across different languages. These findings offer important insights for future expressive speech-to-speech translation systems, particularly for prosody modeling, prosody transfer, and controllable expressive speech generation between multiple languages.


\section{Related Work}

\paragraph{Expressive Speech-to-Speech Translation and Prosody Modeling.}

Traditional speech translation systems commonly relied on cascaded pipelines consisting of automatic speech recognition (ASR), machine translation (MT), and text-to-speech synthesis (TTS) \cite{s2st_direct-jia2019direct}. Prior work noted that such systems may discard paralinguistic information~\cite{prosody_traditional_s2st_discard_prosody-7792638, prosody_ignore-inaguma2023unity, prosody_ignore-kim2024textless, prosody_ignore-lee-etal-2022-direct}. More recent direct speech-to-speech translation (S2ST) approaches, such as Translatotron~2~\cite{s2st_direct-jia2022translatotron}, do not explicitly model cross-lingual prosodic dynamics, and existing evaluations primarily focus on translation quality, naturalness, and speaker similarity rather than fine-grained prosodic preservation. Existing evaluation protocols for these systems primarily focus on speech naturalness, speaker similarity, intelligibility, and translation quality~\cite{mos_analysis-jia2018transfer}, while explicit fine-grained prosodic analysis remains limited. Recent expressive systems such as SeamlessExpressive~\cite{seamless-barrault2023seamless} further model prosodic attributes including pitch, duration, pauses, energy, and expressive conditioning embeddings to improve multilingual speech generation and expressive transfer, but often rely on indirect modeling of prosody or only coarse utterance-level prosodic supervision.

\paragraph{Cross-Lingual Prosody Transfer and Multilingual Dubbing Analysis.}

Studies suggest that current S2ST systems exhibit limited sensitivity to prosodic structure \cite{prosody_study-tsiamas2024speech, prosody_traditional_s2st_discard_prosody-7792638}. Prior work has limited understanding of cross-lingual prosody transfer \cite{prosody_transfer_study-swiatkowski2023cross, prosody_korean_study-zhou2024prosody, avila2023dialog}. A recent empirical study of human dubbing \cite{brannon2023dubbing} suggest that professional dubbing preserves certain expressive and prosodic characteristics across languages. However, these analyses remain relatively coarse-grained, primarily focusing on aggregate timing constraints, speaking rate, lip synchronization, and broad prosodic statistics rather than fine-grained word- or contour-level cross-lingual prosodic correspondence.

\section{Data Processing and Analysis Pipeline}

\begin{figure*}[t]
    \centering
    \includegraphics[width=1.0\textwidth]{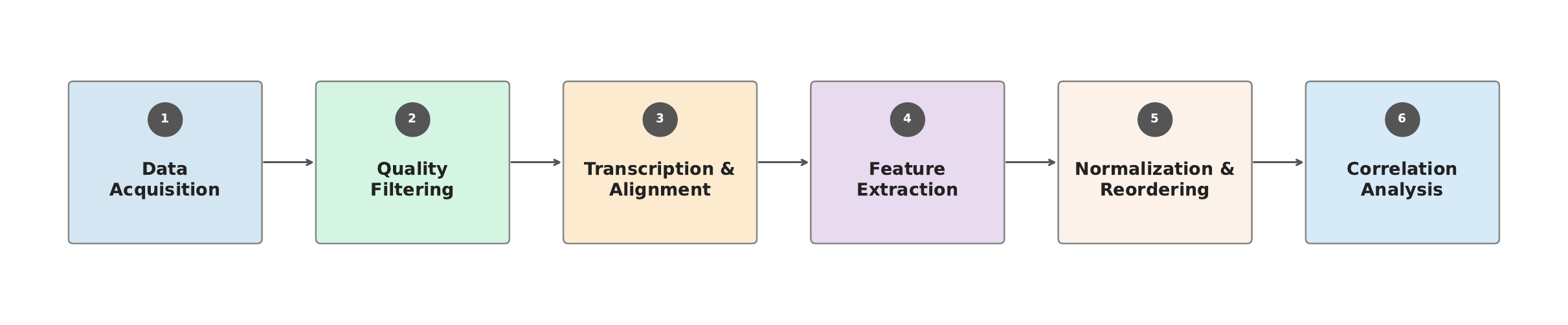}
    \caption{Overview of the cross-lingual prosody analysis pipeline.}
    \label{fig:pipeline}
\end{figure*}

\subsection{Audio Extraction and Quality Filtering}

\paragraph{Data Extraction.}

We analyze multilingual professionally dubbed speech extracted from
television shows. The corpus contains timestamp-aligned source and
dubbed target speech segments across the \texttt{DE--EN}, \texttt{EN--ES},
and \texttt{EN--FR} language pairs. The underlying parallel speech pairs
were automatically mined using the SpeechVecAlign~\cite{speechvecalign-meng2025speech}
pipeline, which aligns semantically corresponding speech segments across
multilingual dubbed audio tracks. Using the provided segment timestamps,
we extract parallel utterance-level speech clips from corresponding
source and dubbed audio tracks for downstream analysis. An utterance is a single conversation turn consisting of a continuous segment of speech produced by one speaker. A parallel pair consists of two utterances: a source utterance and a dubbed target language utterance.

The corpus consists of professionally dubbed entertainment television
episodes spanning multiple genres (e.g., drama, comedy, and action).
The data primarily contains multi-character dialogue rather than
monologue or read speech, providing diverse expressive speech contexts.


\paragraph{Audio Quality Filtering.}
Reliable prosody extraction is highly sensitive to
audio quality. We therefore apply several stages
of filtering before downstream analysis. We first
apply SQUIM objective speech quality
metrics \cite{squim-kumar2023torchaudio-squim} to both source and target
audio. Clips are filtered using thresholds on
SI-SDR, STOI, and PESQ scores to remove noisy,
background-music-heavy, or low-intelligibility
examples.  Table~\ref{tab:dataset_statistics} reports the final number of audio hours retained for each language pair after filtering.

\begin{table}[t]
\centering
\small
\begin{tabular}{llc}
\toprule
\textbf{Stage} & \textbf{Metric} & \textbf{Threshold} \\
\midrule
Audio quality & SI-SDR & $\geq$ 7.5 \\
Audio quality & STOI & $\geq$ 0.91 \\
Audio quality & PESQ & $\geq$ 1.25 \\
\midrule
Diarization & Speakers & $\leq$ 1 \\
\midrule
Semantic filtering & SONAR cosine & $\geq$ 0.63 \\
\bottomrule
\end{tabular}
\caption{
Filtering thresholds used for audio quality,
speaker diarization, and semantic consistency.
}
\label{tab:filter_thresholds}
\vspace{-5mm}
\end{table}

\begin{table}[t]
\centering
\small
\begin{tabular}{lccc}
\toprule
\textbf{Stage} & \textbf{DE-EN} & \textbf{EN-ES} & \textbf{EN-FR} \\
\midrule
Raw pairs               & 100,000 & 100,000 & 100,000 \\
After SQUIM             &  41,610 &  42,974 &  42,045 \\
After diarization       &  28,041 &  28,990 &  29,448 \\
After SONAR             &  20,964 &  22,598 &  20,342 \\
After pitch scoring     &  16,335 &  16,939 &  15,957 \\
\bottomrule
\end{tabular}
\caption{Number of utterance pairs remaining after each
filtering stage. The final row excludes utterance pairs for which valid non-constant pitch contours could not be detected in either the source or target speech.}
\label{tab:filtering_stats}
\vspace{-2mm}
\end{table}

\begin{table}[t]
\centering
\small
\begin{tabular}{lcccc}
\toprule
\textbf{Pair} &
\textbf{Utts.} &
\textbf{Src. hrs} &
\textbf{Tgt. hrs} &
\textbf{Total hrs} \\
\midrule
DE--EN & 16,335 & 18.8 & 17.9 & 36.7 \\
EN--ES & 16,939 & 18.8 & 19.1 & 37.9 \\
EN--FR & 15,957 & 17.6 & 18.1 & 35.7 \\
\bottomrule
\end{tabular}
\caption{Statistics of the final analysis corpus. Hours are computed
over utterances retained for prosodic correlation scoring.}
\label{tab:dataset_statistics}
\vspace{-5mm}
\end{table}

\paragraph{Speaker Diarization.}
Prosodic analysis can be corrupted by overlapping
speech or multiple active speakers. We apply pyannote speaker diarization
\cite{pyannote-bredin2020pyannote} and retain only single-speaker
utterances on both source and target sides.

\paragraph{Semantic Alignment Filtering.}
Parallel speech dataset timestamps occasionally produce
mismatched source-target pairs. To remove
semantically inconsistent examples, we transcribe
both sides of each utterance pair and compute
cross-lingual semantic cosine-similarity using SONAR
embeddings \cite{sonar-duquenne2023sonarsentencelevelmultimodallanguageagnostic}. Pairs below a
similarity threshold are excluded from further
analysis.

\paragraph{Threshold Selection.}
Thresholds were calibrated on a randomly sampled German--English pilot subset consisting of 100 utterance pairs. For each metric,
thresholds were set to the 75th percentile of the
observed score distribution, retaining
approximately the top 25\% of examples by
quality. The same thresholds were subsequently
applied across all language pairs. Tables ~\ref{tab:filter_thresholds}, \ref{tab:filtering_stats}, \ref{tab:dataset_statistics} report the filter thresholds, the number of utterances left after each filtering step and the final dataset size, respectively.

\subsection{Transcription and Word Alignment}

\paragraph{Word-Level Transcription.}
We transcribe all filtered utterances using
\texttt{whisper\_timestamped}\footnote{https://github.com/linto-ai/whisper-timestamped}
\cite{whisper_timestamped-JSSv031i07, whisper_timestamped-lintoai2023whispertimestamped, whisper_timestamped-radford2022robust}, which produces word-level
timestamps for multilingual speech. We use
\texttt{whisper\_timestamped} rather than
standard Whisper because it provides accurate
word-level timing information required for
downstream prosodic alignment. These timestamps are later used to associate prosodic frames with aligned words, as illustrated in Figure~\ref{fig:reordering}.

\paragraph{Text Preprocessing.}
Before alignment, transcripts are normalized using
language-specific preprocessing including
lowercasing, punctuation removal, and contraction
handling.

\paragraph{Bidirectional Word Alignment.}
Word alignment is necessary because semantically corresponding words often appear at different temporal positions across languages due to syntactic reordering and translation variation, making direct left-to-right prosodic comparison less reliable. We perform cross-lingual word alignment using
FastAlign \cite{fastalign-dyer2013simple} with bidirectional
alignment and intersection symmetrization.
Because subtitle utterances are often short and
noisy, we augment training with OPUS TED2013
parallel corpora \cite{opus-TIEDEMANN12.463} to improve
alignment robustness.

\paragraph{Alignment Quality Evaluation.}
To evaluate alignment quality, we measure Alignment Error Rate (AER) on the RWTH German--English gold word alignment benchmark \cite{vilar-etal-2006-aer} derived from Europarl \cite{gold_deen-koehn-2005-europarl}.

FastAlign with OPUS augmentation achieves
82.2\% alignment precision, which
indicates reasonably reliable alignment quality
for downstream prosodic analysis.

\subsection{Prosody Representation}

\paragraph{Pitch and Energy Extraction.}
Pitch (F0) describes the intonation contour of speech and is closely associated with emphasis, emotion, and speaker intent, while energy reflects the intensity or loudness of speech and often correlates with stress, emotion, and emphasis patterns.

\begin{figure*}[t]
    \centering
    \includegraphics[width=0.98\textwidth]{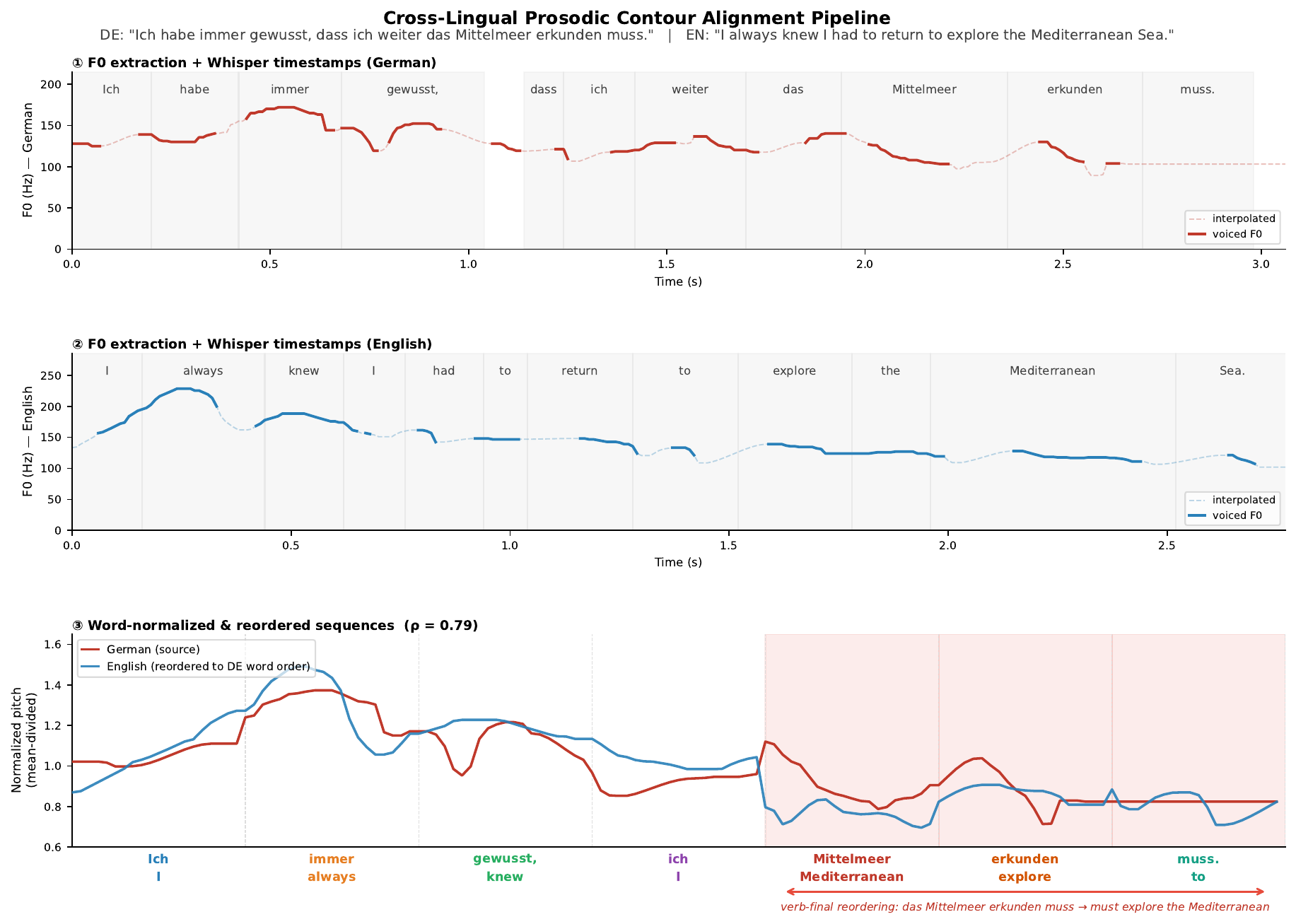}
    \caption{
    Example of the cross-lingual prosody alignment pipeline. Frame-level pitch contours are extracted from source and target utterances using word timestamps and bilingual word alignments. Word-level prosodic segments are grouped into semantically corresponding alignment clusters and reordered prior to correlation analysis.
    }
    \label{fig:reordering}
\end{figure*}

We extract frame-level pitch (F0) using pYAAPT
\cite{pyaapt-kasi2002yet}. Because unvoiced regions occur
inconsistently across languages, we linearly
interpolate missing F0 values to produce
continuous contours. This interpolation strategy is further evaluated against raw F0 contours in Appendix~\ref{app:temporal_robustness}. In addition to pitch, we
extract frame-level energy representations from
the speech signal using librosa \cite{librosa-mcfee2015librosa}. Figure~\ref{fig:reordering} illustrates the resulting frame-level prosodic contours and their alignment with word timestamps.

We observed that pitch extraction produced constant F0 trace on certain audio segments, a failure mode on clean audio where the tracker locks onto a single pitch candidate. We excluded such audio from subsequent calculation. Excluded audio segments did not differ from included segments in utterance length, word count, or alignment ratio, indicating that the exclusion is non-systematic and unlikely to introduce bias.

\paragraph{Word-Level Prosodic Segments.}

For each aligned word pair, we extract the corresponding source and target prosodic segments. Figure~\ref{fig:reordering} illustrates the full contour extraction pipeline, including transcript generation, frame-level pitch extraction, word-level segmentation, and cross-lingual contour alignment.

\paragraph{Cross-Lingual Contour Reordering.}

Direct left-to-right comparison of source and target prosodic contours is unreliable because semantically corresponding content often appears in different word orders across languages. Without reordering, semantically corresponding prosodic events may occur at substantially different temporal locations due to cross-lingual syntactic variation. To address this issue, we use bilingual word alignments to reorder word-level prosodic segments into semantically corresponding sequences prior to utterance-level correlation analysis.

\paragraph{Word-Normalized Contours.}

FastAlign word alignments are grouped into clusters consisting of contiguous source words and their aligned target words, allowing one-to-one, one-to-many, many-to-one, and many-to-many alignments. Each cluster is treated as a single prosodic unit. This produces parallel source and target prosodic sequences of equal length despite structural asymmetries in word alignment.

Each prosodic segment is resampled to a fixed number of frames (\texttt{frames\_per\_word}=20) before concatenation into utterance-level sequences. The robustness of this temporal resolution choice is evaluated in Appendix~\ref{app:temporal_robustness}. This normalization reduces sensitivity to cross-lingual timing differences while preserving local prosodic shape. After concatenation, each utterance-level pitch and energy sequence is normalized by its mean value, reducing absolute speaker-dependent differences and representing contours relative to each speaker's average pitch and energy level.

\subsection{Cross-Lingual Correlation Analysis}

\paragraph{Correlation Metrics.}
For each aligned utterance pair, we compute Spearman Rank Correlation between the source and target prosodic sequences \cite{spearman}. Spearman correlation measures the Pearson correlation between rank-transformed variables and is robust to speaker-dependent absolute pitch differences. We chose Spearman Rank Correlation because prosodic patterns such as emphasis are meaningful in relative terms, while absolute pitch is mostly speaker-dependent.

\paragraph{Shuffled Baseline.}
To verify that observed correlations reflect
genuine cross-lingual prosodic structure rather
than artifacts of smooth contour shapes, we
compute a shuffled baseline by randomly
permuting target prosodic contours within each utterance before
correlation.

\paragraph{Statistical Aggregation.}
We aggregate correlations across utterances using
mean and standard deviation statistics,
and compare results across language pairs and
prosodic feature types.

\paragraph{POS Class Analysis.}
To estimate the contribution of different part-of-speech (POS) categories to cross-lingual prosodic correspondence, we perform a leave-one-group-out analysis. Each alignment cluster is assigned to a POS category based on the dominant POS tag of the English words in that cluster. The POS tags of aligned target-language words are not considered, since cross-lingual POS correspondence is not guaranteed. Assigned clusters are grouped into one of four lexical categories: Nouns (\texttt{NOUN}, \texttt{PROPN}), Verbs (\texttt{VERB}, \texttt{AUX}), Modifiers (\texttt{ADJ}, \texttt{ADV}), and Function words (\texttt{PRON}, \texttt{DET}, \texttt{ADP}, \texttt{CCONJ}, \texttt{SCONJ}, \texttt{PART}). We assign part-of-speech tags using the spaCy \texttt{en\_core\_web\_sm} English pipeline \cite{spaCy-Honnibal_spaCy_Industrial-strength_Natural_2020}.

For each utterance pair, we first compute the baseline zero-lag Spearman correlation using the full word-normalized prosodic contour. We then iteratively remove all prosodic frames associated with a particular POS class from both the source and target sequences and recompute the correlation on the remaining contour. The effect of a POS class is defined as
\[
\Delta_g = r_{\mathrm{full}} - r_{\mathrm{without}\ g},
\]
where positive values indicate that removing the POS class decreases correlation, suggesting that the group affects the cross-lingual prosodic preservation positively.

The leave-one-group-out design preserves the majority of the utterance-level contour structure while selectively ablating a specific POS category. This reduces instability arising from recomputing rank correlations on short or fragmented subsequences and enables more reliable estimation of each group's effect on overall prosodic correspondence.

\subsection{Temporal Feature Analysis}

In addition to fine-grained contour-level prosody analysis, the availability of large-scale high quality filtered multilingual dubbing data enables analysis of broader utterance-level temporal correspondence across languages. We therefore perform a coarse temporal analysis using utterance duration and total phoneme count.

\paragraph{Alignment Filtering.}
To ensure consistent segmentation, we exclude aligned speech segments
for which the source and target yielded differing
number of sentences for an accurate utterance-by-utterance analysis.

\paragraph{Phonetic and Acoustic Analysis.}
For each utterance pair, we extract two features: the total phoneme count from the ASR transcripts using phonemizer \cite{bernard2021phonemizer} and the utterance duration from word-level timestamps.

\paragraph{Correlation Metrics.}
We calculate Spearman rank correlation \cite{spearman} between the number of phonemes and duration observations from the source and target utterances. 

\section{Prosody Analysis Results}

\begin{figure*}[t!]
    \centering
    \includegraphics[width=0.95\textwidth]{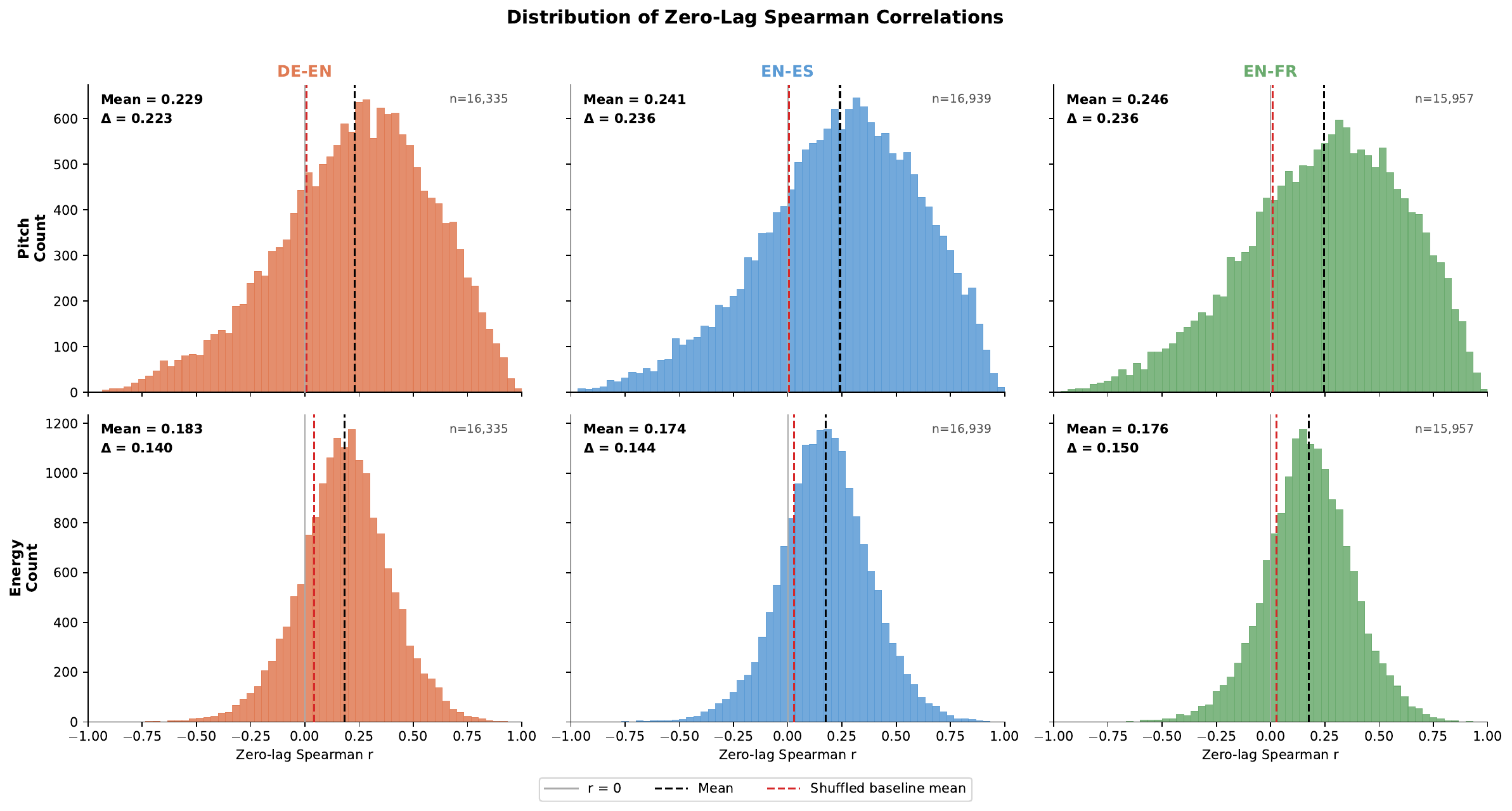}
    \caption{
Distribution of utterance-level zero-lag Spearman correlations for pitch and energy across aligned source--target utterance pairs.
}
    \label{fig:pitch_corr_distribution}
\end{figure*}

\subsection{Prosodic Correlation Analysis}
\label{sec:prosody_results}

Table~\ref{tab:prosody_corr} summarizes utterance-level cross-lingual prosodic correlations across all examined language pairs. We observe consistently positive correlations for both pitch and energy in all language pairs, substantially exceeding the shuffled baseline. Pitch correlations are consistently stronger than energy correlations across all examined language pairs, with mean pitch correlations ranging from 0.229 to 0.246, compared to 0.174 to 0.183 for energy.

\begin{table}[t]
\centering
\small
\setlength{\tabcolsep}{4pt}
\begin{tabular}{llccc}
\toprule
\textbf{Feature} &
\textbf{Pair} &
\textbf{Mean $r$} &
\textbf{95\% CI} &
\textbf{Shuffled} \\
\midrule
Pitch  & DE--EN & 0.229 & [0.223, 0.236] & 0.007 \\
       & EN--ES & 0.241 & [0.235, 0.247] & 0.005 \\
       & EN--FR & 0.246 & [0.239, 0.253] & 0.010 \\
\midrule
Energy & DE--EN & 0.183 & [0.179, 0.187] & 0.043 \\
       & EN--ES & 0.174 & [0.170, 0.178] & 0.030 \\
       & EN--FR & 0.176 & [0.172, 0.179] & 0.026 \\
\bottomrule
\end{tabular}
\caption{Mean utterance-level Spearman correlation between aligned
source and target prosodic contours. Confidence intervals are computed
using episode-clustered bootstrap resampling. The shuffled baseline
randomizes contours within each utterance. All aligned
correlations significantly exceed the shuffled baseline
(paired Wilcoxon, $p < 10^{-80}$).}
\label{tab:prosody_corr}
\end{table}

Figure~\ref{fig:pitch_corr_distribution} further visualizes the distribution of utterance-level correlations. Across all language pairs, the observed distributions are clearly shifted toward positive correlation values relative to the shuffled baseline, indicating systematic cross-lingual prosodic correspondence. Importantly, this positive shift is not driven solely by a small number of highly correlated examples, but instead reflects a broad tendency toward positive correlation across utterances. Pitch distributions additionally exhibit broader variance and stronger positive tails than energy, suggesting greater variability in intonational preservation across utterances.

Although the absolute magnitude of the correlations remains moderate, this is expected given the challenging nature of cross-lingual prosody comparison. Unlike within-speaker or within-language settings, dubbed speech introduces substantial variability from translation divergence, timing shifts, phonetic inventory differences, independent voice acting choices, and speaker identity variation. Moreover, frame-level pitch and energy are highly detailed signals containing both meaningful prosodic variation and local noise. Under such conditions, even moderate positive correlations substantially above the shuffled baseline indicate non-trivial cross-lingual preservation of prosodic structure.


We additionally observe minor differences across language pairs. English–Spanish and English–French exhibit slightly stronger prosodic correspondence than German–English in terms of pitch, while German–English is slightly stronger for energy, potentially reflecting differences in syntactic structure, rhythmic organization, or dubbing conventions across language pairs. Overall the correlation values are similar.
Nevertheless, all examined language pairs consistently display evidence of prosodic similarity above the randomized baseline. Taken together, these findings suggest that prosodic organization is not entirely language-specific, but instead contains partially shared cross-lingual structure that persists across translation and professional dubbing.

We further examine the statistical robustness of these correlations in Appendix~\ref{app:additional_prosodic_correlation}. Pairwise comparisons between language pairs and paired comparisons between pitch and energy show that, while some differences are statistically significant due to the large sample size, the effect sizes are negligible across language pairs, whereas pitch consistently exhibits stronger correspondence than energy.

\begin{table}[t]
\centering
\small
\begin{tabular}{llccc}
\toprule
Feature & Class & DE--EN & EN--ES & EN--FR \\
\midrule
\multirow{4}{*}{Pitch}
& Nouns     & \textbf{+0.0228} & \textbf{+0.0208} & \textbf{+0.0203} \\
& Verbs     & +0.0034 & +0.0156 & +0.0045 \\
& Modifiers & +0.0083 & +0.0057 & +0.0055 \\
& Function  & -0.0065 & +0.0108 & +0.0132 \\
\midrule
\multirow{4}{*}{Energy}
& Nouns     & \textbf{+0.0355} & \textbf{+0.0396} & \textbf{+0.0402} \\
& Verbs     & -0.0041 & -0.0064 & -0.0128 \\
& Modifiers & +0.0027 & +0.0062 & +0.0052 \\
& Function  & \textbf{-0.0324} & \textbf{-0.0203} & \textbf{-0.0204} \\
\bottomrule
\end{tabular}
\caption{
Leave-one-group-out POS class analysis. Positive values indicate that the POS class has a positive effect on cross-lingual prosodic correspondence.
}
\label{tab:lexical_contrib}
\end{table}

\begin{figure*}[t!]
    \centering
    \includegraphics[width=0.95\textwidth]{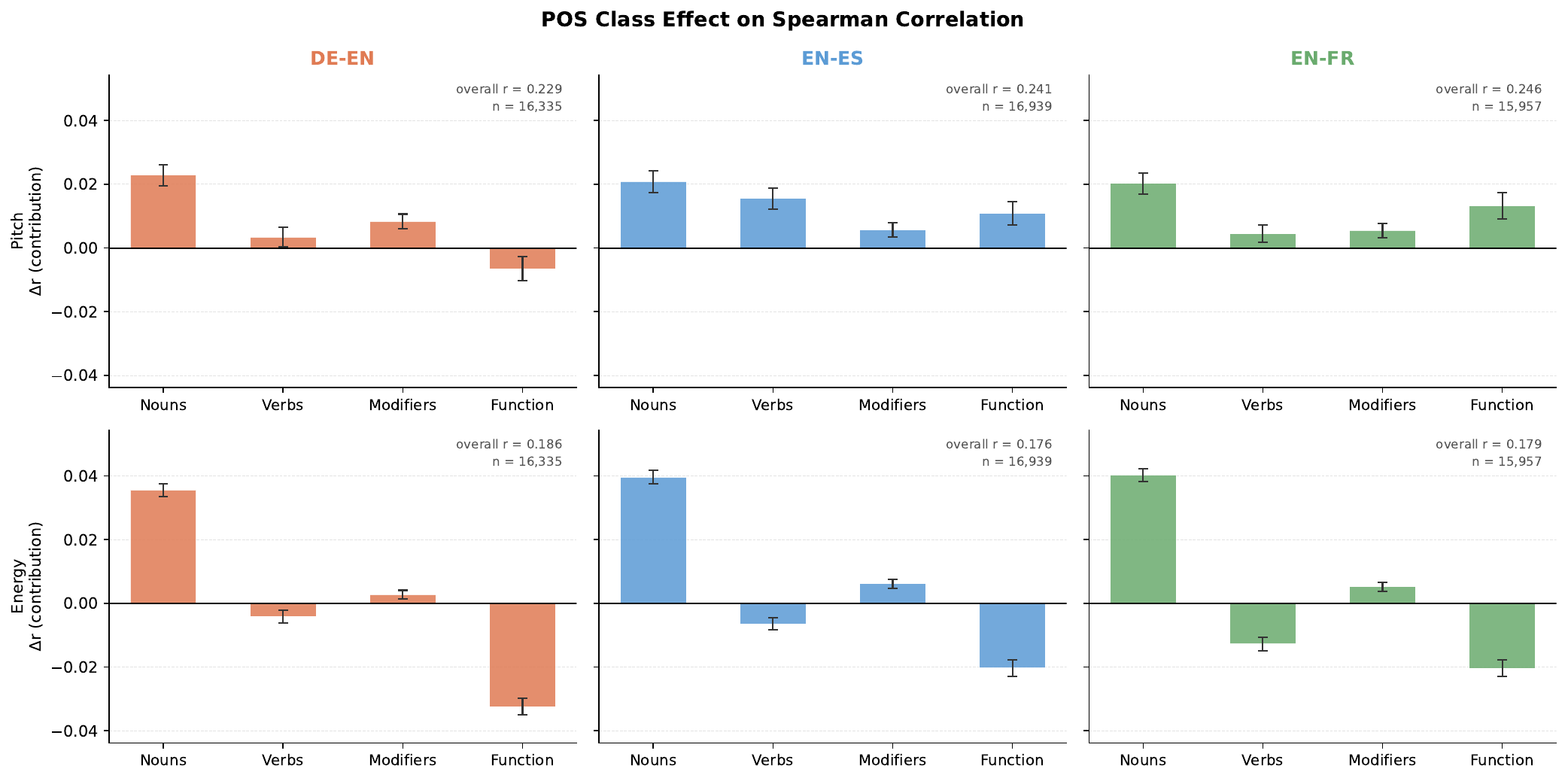}
    \caption{
Leave-one-group-out POS class analysis across language pairs and prosodic dimensions. Bars denote the change in utterance-level Spearman correlation after removing all prosodic frames associated with a POS category.
Positive values indicate that the lexical class has a positive effect on cross-lingual prosodic correspondence. Error bars denote 95\% confidence intervals on the mean effect, computed as $1.96 \cdot (\sigma / \sqrt{n})$, where $\sigma$ is the standard deviation of per-utterance values.
}
    \label{fig:pos_contribution}
    \vspace{-3mm}
\end{figure*}

\subsection{POS Class Analysis}

Figure~\ref{fig:pos_contribution} and Table~\ref{tab:lexical_contrib}
summarize the change in utterance-level Spearman correlation after
removing all prosodic frames associated with a POS class.

Across both pitch and energy, nouns consistently show the strongest
positive effect across all language pairs. Removing noun-associated
prosodic frames produces the largest reduction in measured
cross-lingual prosodic correspondence, suggesting that noun-associated
regions affect the observed prosodic similarity the most.
This effect is particularly strong for energy, where noun removal
decreases correlation by approximately 0.036--0.040 across language
pairs.

In contrast, function words exhibit substantially different behavior
across prosodic dimensions. For energy, removing function words
consistently increases correlation across all language pairs, suggesting
that function-word-associated regions introduce greater variability in
cross-lingual energy correspondence. Verbs exhibit similarly weak or
slightly negative effects, while modifiers remain near zero.

Pitch exhibits a more heterogeneous pattern than energy. While
noun-associated regions remain positively associated with overall
correlation, pitch effects are generally smaller and
less consistent across POS categories. Function words affect
positively for EN--ES and EN--FR but negatively for DE--EN, suggesting
that POS-category effects on pitch correspondence vary across
language pairs. Pitch effects also exhibit larger variance across
utterances than energy effects, indicating greater variability in
intonational correspondence.

Overall, these results show that measured cross-lingual prosodic
correspondence is not distributed uniformly across POS categories.
Noun-associated regions affect most strongly to the observed
correspondence, while function-word-associated regions, particularly
for energy, show less stable patterns across languages. Additional
statistical analyses of these POS-class effects, including bootstrap
confidence intervals and paired comparisons, are provided in
Appendix~\ref{app:lexical_results}.

\subsection{Temporal Feature Analysis}

Table~\ref{tab:temporal_corr} presents utterance-level cross-lingual temporal correlations across language pairs. Both utterance duration and total phoneme count exhibit consistently strong positive correlations between source and target speech, which is expected given the temporal synchronization constraints of professional dubbing. We further analyze phoneme count correspondence to investigate cross-lingual temporal isometry, since substantial differences in phoneme realization would likely manifest as variations in speaking rate across languages.

Duration correlations are slightly stronger than phoneme-count correlations across all language pairs, indicating that broad temporal organization is highly consistent between source and target speech. The consistently strong correlations across multiple language pairs suggest that coarse timing structure is robustly preserved across translation despite differences in phonetic inventory and syntactic structure.

These findings suggest that temporal isometry between these languages is sufficiently high that dubbing actors do not need to drastically alter speaking rate to satisfy timing constraints.

\begin{table}[t!]
\centering
\small
\setlength{\tabcolsep}{3.5pt}
\begin{tabular}{llrrrr}
\toprule
Pair & Feature & Corr.\\
\midrule
DE--EN & Phonemes    & 0.885 \\
EN--ES & Phonemes    & 0.875 \\
EN--FR & Phonemes    & 0.871 \\
\midrule
DE--EN & Duration    & 0.891 \\
EN--ES & Duration    & 0.890 \\
EN--FR & Duration    & 0.884 \\
\bottomrule
\end{tabular}
\caption{
Utterance-level correlations between source and target temporal features (total number of phonemes, duration) for each language pair.
}
\label{tab:temporal_corr}
\end{table}

\section{Conclusion}

In this work, we presented the first large-scale fine-grained analysis of cross-lingual prosodic similarity using professionally dubbed multilingual speech. Using alignment-aware word-level contour analysis across German–English, English–Spanish, and English–French language pairs, we investigated how pitch, energy, and temporal speech patterns are preserved across translation.

Our results show consistent positive cross-lingual correlations for both pitch and energy substantially above shuffled baselines, indicating that prosodic structure is partially preserved across languages despite differences in syntax, phonetics, speaker identity, and dubbing style. Pitch generally exhibits stronger but more variable correspondence than energy, suggesting that intonational structure may be preserved at a broader utterance-level scale while energy patterns are more locally dependent on lexical realization.

We further analyzed the contribution of different POS categories to
cross-lingual prosodic correspondence. Noun-associated regions consistently
showed the strongest positive contribution across language pairs, while
function-word-associated regions exhibited less stable patterns,
particularly for energy. These findings suggest that measured prosodic
correspondence is not uniformly distributed across POS categories,
and that regions associated with content-bearing words may contribute
more strongly to the shared prosodic patterns observed across languages.

Finally, we observed consistently strong cross-lingual correspondence in coarse utterance-level temporal features, with duration exhibiting slightly stronger correlations than total phoneme count across all language pairs. Compared to fine-grained pitch and energy contours, these substantially larger correlations are expected, since localized contour alignment is inherently more sensitive to variation in intonation, speaker style, and translation reordering.

Our findings provide new empirical insights into cross-lingual prosody transfer and highlight the importance of alignment-aware analysis for future expressive speech translation systems. We hope this work motivates future research on prosody-aware multilingual speech modeling, controllable expressive translation, and cross-lingual speech generation.

\section*{Limitations}

This work has several limitations. First, professionally dubbed speech only approximates prosody transfer, as actors may alter timing, emphasis, or intonation for fluency and synchronization. Second, we focus on pitch, energy, and coarse temporal features, leaving pauses, rhythm, and voice quality unexplored. Third, automatic transcription and bilingual word alignment may introduce noise despite filtering. Our experiments are also limited to three European language pairs and may not generalize to typologically distant or tonal languages. Finally, the multilingual dubbing corpus cannot be publicly released due to licensing restrictions. We plan to release the full analysis pipeline to support reproduction on compatible datasets.

\section*{Acknowledgments}
This work was supported by the National Research Foundation, Singapore, and the Ministry of Digital Development and Information under its Singapore Global AI Visiting Professorship (AIVP-2026-010), and in part by the National Science Foundation (NSF) under CAREER Award IIS-2533652. Any opinions, findings and conclusions or recommendations expressed in this material are those of the author(s) and do not reflect the views of National Research Foundation, Singapore and the Ministry of Digital Development and Information.

\paragraph{AI Disclosure}

AI assistants were used for language editing, writing refinement, brainstorming, and minor coding assistance. All technical decisions, experiments, analyses, and final manuscript content were reviewed and verified by the authors.


\bibliography{anthology,custom}

@inproceedings{prosody_ignore-inaguma2023unity,
    title = "{U}nit{Y}: Two-pass Direct Speech-to-speech Translation with Discrete Units",
    author = "Inaguma, Hirofumi  and
      Popuri, Sravya  and
      Kulikov, Ilia  and
      Chen, Peng-Jen  and
      Wang, Changhan  and
      Chung, Yu-An  and
      Tang, Yun  and
      Lee, Ann  and
      Watanabe, Shinji  and
      Pino, Juan",
    editor = "Rogers, Anna  and
      Boyd-Graber, Jordan  and
      Okazaki, Naoaki",
    booktitle = "Proceedings of the 61st Annual Meeting of the Association for Computational Linguistics (Volume 1: Long Papers)",
    month = jul,
    year = "2023",
    address = "Toronto, Canada",
    publisher = "Association for Computational Linguistics",
    url = "https://aclanthology.org/2023.acl-long.872/",
    doi = "10.18653/v1/2023.acl-long.872",
    pages = "15655--15680"
}

@inproceedings{prosody_ignore-lee-etal-2022-direct,
    title = "Direct Speech-to-Speech Translation With Discrete Units",
    author = "Lee, Ann  and
      Chen, Peng-Jen  and
      Wang, Changhan  and
      Gu, Jiatao  and
      Popuri, Sravya  and
      Ma, Xutai  and
      Polyak, Adam  and
      Adi, Yossi  and
      He, Qing  and
      Tang, Yun  and
      Pino, Juan  and
      Hsu, Wei-Ning",
    editor = "Muresan, Smaranda  and
      Nakov, Preslav  and
      Villavicencio, Aline",
    booktitle = "Proceedings of the 60th Annual Meeting of the Association for Computational Linguistics (Volume 1: Long Papers)",
    month = may,
    year = "2022",
    address = "Dublin, Ireland",
    publisher = "Association for Computational Linguistics",
    url = "https://aclanthology.org/2022.acl-long.235/",
    doi = "10.18653/v1/2022.acl-long.235",
    pages = "3327--3339"
}

@inproceedings{prosody_korean_study-zhou2024prosody,
    title = "Prosody in Cascade and Direct Speech-to-Text Translation: a case study on {K}orean Wh-Phrases",
    author = "Zhou, Giulio  and
      Lam, Tsz Kin  and
      Birch, Alexandra  and
      Haddow, Barry",
    editor = "Graham, Yvette  and
      Purver, Matthew",
    booktitle = "Findings of the Association for Computational Linguistics: EACL 2024",
    month = mar,
    year = "2024",
    address = "St. Julian{'}s, Malta",
    publisher = "Association for Computational Linguistics",
    url = "https://aclanthology.org/2024.findings-eacl.46/",
    doi = "10.18653/v1/2024.findings-eacl.46",
    pages = "674--683"
}

@article{brannon2023dubbing,
   title={Dubbing in Practice: A Large Scale Study of Human Localization With Insights for Automatic Dubbing},
   volume={11},
   ISSN={2307-387X},
   url={http://dx.doi.org/10.1162/tacl_a_00551},
   DOI={10.1162/tacl_a_00551},
   journal={Transactions of the Association for Computational Linguistics},
   publisher={MIT Press},
   author={Brannon, William and Virkar, Yogesh and Thompson, Brian},
   year={2023},
   pages={419–435} }

@inproceedings{speechvecalign-meng2025speech,
    title = "Speech Vecalign: an Embedding-based Method for Aligning Parallel Speech Documents",
    author = "Meng, Chutong  and
      Koehn, Philipp",
    editor = "Christodoulopoulos, Christos  and
      Chakraborty, Tanmoy  and
      Rose, Carolyn  and
      Peng, Violet",
    booktitle = "Proceedings of the 2025 Conference on Empirical Methods in Natural Language Processing",
    month = nov,
    year = "2025",
    address = "Suzhou, China",
    publisher = "Association for Computational Linguistics",
    url = "https://aclanthology.org/2025.emnlp-main.833/",
    doi = "10.18653/v1/2025.emnlp-main.833",
    pages = "16478--16494",
    ISBN = "979-8-89176-332-6"
}

@inproceedings{fastalign-dyer2013simple,
    title = "A Simple, Fast, and Effective Reparameterization of {IBM} Model 2",
    author = "Dyer, Chris  and
      Chahuneau, Victor  and
      Smith, Noah A.",
    editor = "Vanderwende, Lucy  and
      Daum{\'e} III, Hal  and
      Kirchhoff, Katrin",
    booktitle = "Proceedings of the 2013 Conference of the North {A}merican Chapter of the Association for Computational Linguistics: Human Language Technologies",
    month = jun,
    year = "2013",
    address = "Atlanta, Georgia",
    publisher = "Association for Computational Linguistics",
    url = "https://aclanthology.org/N13-1073/",
    pages = "644--648"
}

@inproceedings{gold_deen-koehn-2005-europarl,
    title = "{E}uroparl: A Parallel Corpus for Statistical Machine Translation",
    author = "Koehn, Philipp",
    booktitle = "Proceedings of Machine Translation Summit X: Papers",
    month = sep # " 13-15",
    year = "2005",
    address = "Phuket, Thailand",
    url = "https://aclanthology.org/2005.mtsummit-papers.11/",
    pages = "79--86"
}

@inproceedings{french_ssml-ouali2025improving,
    title = "Improving {F}rench Synthetic Speech Quality via {SSML} Prosody Control",
    author = "Ouali, Nassima Ould  and
      Sani, Awais Hussain  and
      Bueno, Ruben  and
      Dauvet, Jonah  and
      Horstmann, Tim Luka  and
      Moulines, Eric",
    editor = "Abbas, Mourad  and
      Yousef, Tariq  and
      Galke, Lukas",
    booktitle = "Proceedings of the 8th International Conference on Natural Language and Speech Processing (ICNLSP-2025)",
    month = aug,
    year = "2025",
    address = "Southern Denmark University, Odense, Denmark",
    publisher = "Association for Computational Linguistics",
    url = "https://aclanthology.org/2025.icnlsp-1.30/",
    pages = "302--314"
}

@inproceedings{prosody_study-tsiamas2024speech,
    title = "Speech Is More than Words: Do Speech-to-Text Translation Systems Leverage Prosody?",
    author = "Tsiamas, Ioannis  and
      Sperber, Matthias  and
      Finch, Andrew  and
      Garg, Sarthak",
    editor = "Haddow, Barry  and
      Kocmi, Tom  and
      Koehn, Philipp  and
      Monz, Christof",
    booktitle = "Proceedings of the Ninth Conference on Machine Translation",
    month = nov,
    year = "2024",
    address = "Miami, Florida, USA",
    publisher = "Association for Computational Linguistics",
    url = "https://aclanthology.org/2024.wmt-1.119/",
    doi = "10.18653/v1/2024.wmt-1.119",
    pages = "1235--1257"
}

@inproceedings{opus-TIEDEMANN12.463,
    title = "Parallel Data, Tools and Interfaces in {OPUS}",
    author = {Tiedemann, J{\"o}rg},
    editor = "Calzolari, Nicoletta  and
      Choukri, Khalid  and
      Declerck, Thierry  and
      Do{\u{g}}an, Mehmet U{\u{g}}ur  and
      Maegaard, Bente  and
      Mariani, Joseph  and
      Moreno, Asuncion  and
      Odijk, Jan  and
      Piperidis, Stelios",
    booktitle = "Proceedings of the Eighth International Conference on Language Resources and Evaluation ({LREC}'12)",
    month = may,
    year = "2012",
    address = "Istanbul, Turkey",
    publisher = "European Language Resources Association (ELRA)",
    url = "https://aclanthology.org/L12-1246/",
    pages = "2214--2218"
}

@inproceedings{vilar-etal-2006-aer,
    title = "{AER}: do we need to ``improve'' our alignments?",
    author = "Vilar, David  and
      Popovic, Maja  and
      Ney, Hermann",
    booktitle = "Proceedings of the Third International Workshop on Spoken Language Translation: Papers",
    month = nov # " 27-28",
    year = "2006",
    address = "Kyoto, Japan",
    url = "https://aclanthology.org/2006.iwslt-papers.7/"
}

@article{spearman,
 ISSN = {00029556},
 URL = {http://www.jstor.org/stable/1412159},
 author = {C. Spearman},
 journal = {The American Journal of Psychology},
 number = {1},
 pages = {72--101},
 publisher = {University of Illinois Press},
 title = {The Proof and Measurement of Association between Two Things},
 urldate = {2026-05-25},
 volume = {15},
 year = {1904}
}

@incollection{prosody_importance-10.1093/oxfordhb/9780198791768.013.30,
    author = {Tonhauser, Judith},
    editor = {Cummins, Chris and Katsos, Napoleon},
    isbn = {9780198791768},
    title = {Prosody and Meaning},
    booktitle = {The Oxford Handbook of Experimental Semantics and Pragmatics},
    publisher = {Oxford University Press},
    year = {2019},
    month = {03},
    doi = {10.1093/oxfordhb/9780198791768.013.30},
    url = {https://doi.org/10.1093/oxfordhb/9780198791768.013.30},
    eprint = {https://academic.oup.com/book/0/chapter/335410307/chapter-ag-pdf/45911661/book_38644_section_335410307.ag.pdf},
}

@article{prosody_importance-article,
author = {Cutler, Anne and Dahan, Delphine and Donselaar, Wilma},
year = {1997},
month = {04},
pages = {141-201},
title = {Prosody in the Comprehension of Spoken Language: A Literature Review},
volume = {40 ( Pt 2)},
journal = {Language and speech},
doi = {10.1177/002383099704000203}
}

@book{prosody_importance-Gussenhoven_2004, place={Cambridge}, series={Research Surveys in Linguistics}, title={The Phonology of Tone and Intonation}, publisher={Cambridge University Press}, author={Gussenhoven, Carlos}, year={2004}, collection={Research Surveys in Linguistics}}

@article{prosody_ignore-kim2024textless,
  title={Textless unit-to-unit training for many-to-many multilingual speech-to-speech translation},
  author={Kim, Minsu and Choi, Jeongsoo and Kim, Dahun and Ro, Yong Man},
  journal={IEEE/ACM Transactions on Audio, Speech, and Language Processing},
  volume={32},
  pages={3934--3946},
  year={2024},
  publisher={IEEE}
}

@InProceedings{s2st_direct-jia2022translatotron,
  title = 	 {Translatotron 2: High-quality direct speech-to-speech translation with voice preservation},
  author =       {Jia, Ye and Ramanovich, Michelle Tadmor and Remez, Tal and Pomerantz, Roi},
  booktitle = 	 {Proceedings of the 39th International Conference on Machine Learning},
  pages = 	 {10120--10134},
  year = 	 {2022},
  editor = 	 {Chaudhuri, Kamalika and Jegelka, Stefanie and Song, Le and Szepesvari, Csaba and Niu, Gang and Sabato, Sivan},
  volume = 	 {162},
  series = 	 {Proceedings of Machine Learning Research},
  month = 	 {17--23 Jul},
  publisher =    {PMLR},
  url = 	 {https://proceedings.mlr.press/v162/jia22b.html}
}

@inproceedings{s2st_direct-jia2019direct,
  title     = {{Direct Speech-to-Speech Translation with a Sequence-to-Sequence Model}},
  author    = {Ye Jia and Ron J. Weiss and Fadi Biadsy and Wolfgang Macherey and Melvin Johnson and Zhifeng Chen and Yonghui Wu},
  year      = {2019},
  booktitle = {{Interspeech 2019}},
  pages     = {1123--1127},
  doi       = {10.21437/Interspeech.2019-1951},
  issn      = {2958-1796},
}

@inproceedings{mos_analysis-jia2018transfer,
author = {Jia, Ye and Zhang, Yu and Weiss, Ron J. and Wang, Quan and Shen, Jonathan and Ren, Fei and Chen, Zhifeng and Nguyen, Patrick and Pang, Ruoming and Moreno, Ignacio Lopez and Wu, Yonghui},
title = {Transfer learning from speaker verification to multispeaker text-to-speech synthesis},
year = {2018},
publisher = {Curran Associates Inc.},
address = {Red Hook, NY, USA},
booktitle = {Proceedings of the 32nd International Conference on Neural Information Processing Systems},
pages = {4485–4495},
numpages = {11},
location = {Montr{\'e}al, Canada},
series = {NIPS'18}
}

@misc{seamless-barrault2023seamless,
      title={Seamless: Multilingual Expressive and Streaming Speech Translation}, 
      author={{Seamless Communication} and Loïc Barrault and Yu-An Chung and Mariano Coria Meglioli and David Dale and Ning Dong and Mark Duppenthaler and Paul-Ambroise Duquenne and Brian Ellis and others},
      year={2023},
      eprint={2312.05187},
      archivePrefix={arXiv},
      primaryClass={cs.CL},
      url={https://arxiv.org/abs/2312.05187}, 
}

@article{prosody_transfer_study-swiatkowski2023cross,
  title={Cross-lingual prosody transfer for expressive machine dubbing},
  author={Swiatkowski, Jakub and Wang, Duo and Babianski, Mikolaj and Tobing, Patrick Lumban and Vipperla, Ravichander and Pollet, Vincent},
  journal={arXiv preprint arXiv:2306.11658},
  year={2023}
}

@ARTICLE{prosody_traditional_s2st_discard_prosody-7792638,
  author={Do, Quoc Truong and Toda, Tomoki and Neubig, Graham and Sakti, Sakriani and Nakamura, Satoshi},
  journal={IEEE/ACM Transactions on Audio, Speech, and Language Processing}, 
  title={Preserving Word-Level Emphasis in Speech-to-Speech Translation}, 
  year={2017},
  volume={25},
  number={3},
  pages={544-556},
  doi={10.1109/TASLP.2016.2643280}}

@inproceedings{avila2023dialog,
  title     = {{Towards Cross-Language Prosody Transfer for Dialog}},
  author    = {Jonathan E. Avila and Nigel G. Ward},
  year      = {2023},
  booktitle = {{Interspeech 2023}},
  pages     = {2143--2147},
  doi       = {10.21437/Interspeech.2023-1152},
  issn      = {2958-1796},
}

@INPROCEEDINGS{rasim-passt,
  author={Ulgen, Ismail Rasim and Liu, Nancy and Sadoughi, Najmeh and Yanamandra, Abhishek and Jain, Abhinav and Liu, Zhu and Bhat, Vimal},
  booktitle={ICASSP 2026 - 2026 IEEE International Conference on Acoustics, Speech and Signal Processing (ICASSP)}, 
  title={Direct Transfer of Prosody in Speech-to-speech Translation using Disentangled Speech Tokens}, 
  year={2026},
  volume={},
  number={},
  pages={18372-18376},
  doi={10.1109/ICASSP55912.2026.11464610}}

@inproceedings{pyannote-bredin2020pyannote,
  title={Pyannote. audio: neural building blocks for speaker diarization},
  author={Bredin, Herv{\'e} and Yin, Ruiqing and Coria, Juan Manuel and Gelly, Gregory and Korshunov, Pavel and Lavechin, Marvin and Fustes, Diego and Titeux, Hadrien and Bouaziz, Wassim and Gill, Marie-Philippe},
  booktitle={ICASSP 2020-2020 IEEE International conference on acoustics, speech and signal processing (ICASSP)},
  pages={7124--7128},
  year={2020},
  organization={IEEE}
}

@INPROCEEDINGS{squim-kumar2023torchaudio-squim,
  author={Kumar, Anurag and Tan, Ke and Ni, Zhaoheng and Manocha, Pranay and Zhang, Xiaohui and Henderson, Ethan and Xu, Buye},
  booktitle={ICASSP 2023 - 2023 IEEE International Conference on Acoustics, Speech and Signal Processing (ICASSP)}, 
  title={Torchaudio-Squim: Reference-Less Speech Quality and Intelligibility Measures in Torchaudio}, 
  year={2023},
  volume={},
  number={},
  pages={1-5},
  doi={10.1109/ICASSP49357.2023.10096680}}

@misc{sonar-duquenne2023sonarsentencelevelmultimodallanguageagnostic,
      title={SONAR: Sentence-Level Multimodal and Language-Agnostic Representations}, 
      author={Paul-Ambroise Duquenne and Holger Schwenk and Benoît Sagot},
      year={2023},
      eprint={2308.11466},
      archivePrefix={arXiv},
      primaryClass={cs.CL},
      url={https://arxiv.org/abs/2308.11466}, 
}

@misc{whisper_timestamped-lintoai2023whispertimestamped,
  title={whisper-timestamped},
  author={Louradour, J{\'e}r{\^o}me},
  journal={GitHub repository},
  year={2023},
  publisher={GitHub},
  howpublished = {\url{https://github.com/linto-ai/whisper-timestamped}}
}

@InProceedings{whisper_timestamped-radford2022robust,
  title = 	 {Robust Speech Recognition via Large-Scale Weak Supervision},
  author =       {Radford, Alec and Kim, Jong Wook and Xu, Tao and Brockman, Greg and Mcleavey, Christine and Sutskever, Ilya},
  booktitle = 	 {Proceedings of the 40th International Conference on Machine Learning},
  pages = 	 {28492--28518},
  year = 	 {2023},
  editor = 	 {Krause, Andreas and Brunskill, Emma and Cho, Kyunghyun and Engelhardt, Barbara and Sabato, Sivan and Scarlett, Jonathan},
  volume = 	 {202},
  series = 	 {Proceedings of Machine Learning Research},
  month = 	 {23--29 Jul},
  publisher =    {PMLR},
  url = 	 {https://proceedings.mlr.press/v202/radford23a.html}
}

@article{whisper_timestamped-JSSv031i07,
  title={Computing and Visualizing Dynamic Time Warping Alignments in R: The dtw Package},
  author={Giorgino, Toni},
  journal={Journal of Statistical Software},
  year={2009},
  volume={31},
  number={7},
  doi={10.18637/jss.v031.i07}
}

@INPROCEEDINGS{pyaapt-kasi2002yet,
  author={Kasi, Kavita and Zahorian, Stephen A.},
  booktitle={2002 IEEE International Conference on Acoustics, Speech, and Signal Processing}, 
  title={Yet Another Algorithm for Pitch Tracking}, 
  year={2002},
  volume={1},
  number={},
  pages={I-361-I-364},
  doi={10.1109/ICASSP.2002.5743729}}

@article{librosa-mcfee2015librosa,
  author = {McFee, Brian and Raffel, Colin and Liang, Dawen and Ellis, Daniel P.W. and McVicar, Matt and Battenberg, Eric and Nieto, Oriol},
  title = {librosa: Audio and Music Signal Analysis in Python},
  journal = {SciPy 2015},
  year = {2015},
  doi = {10.25080/Majora-7b98e3ed-003},
  url = {https://doi.org/10.25080/Majora-7b98e3ed-003}
}

@article{spaCy-Honnibal_spaCy_Industrial-strength_Natural_2020,
author = {Honnibal, Matthew and Montani, Ines and Van Landeghem, Sofie and Boyd, Adriane},
doi = {10.5281/zenodo.1212303},
title = {{spaCy: Industrial-strength Natural Language Processing in Python}},
year = {2020}
}

@article{bernard2021phonemizer,
  title={Phonemizer: Text to phones transcription for multiple languages in python},
  author={Bernard, Mathieu and Titeux, Hadrien},
  journal={Journal of Open Source Software},
  volume={6},
  number={68},
  pages={3958},
  year={2021}
}

@INPROCEEDINGS{gendertool,
  author={Doukhan, David and Carrive, Jean and Vallet, Felicien and Larcher, Anthony and Meignier, Sylvain},
  booktitle={2018 IEEE International Conference on Acoustics, Speech and Signal Processing (ICASSP)}, 
  title={An Open-Source Speaker Gender Detection Framework for Monitoring Gender Equality}, 
  year={2018},
  volume={},
  number={},
  pages={5214-5218},
  doi={10.1109/ICASSP.2018.8461471}}

\appendix

\setcounter{topnumber}{3}
\setcounter{bottomnumber}{2}
\setcounter{totalnumber}{5}
\setcounter{dbltopnumber}{4}

\renewcommand{\topfraction}{0.9}
\renewcommand{\bottomfraction}{0.6}
\renewcommand{\dbltopfraction}{0.9}
\renewcommand{\textfraction}{0.07}
\renewcommand{\floatpagefraction}{0.75}
\renewcommand{\dblfloatpagefraction}{0.75}

\section{Additional Alignment Evaluation}

To evaluate the quality of the word alignment component, we compare FastAlign symmetrization heuristics on the German--English gold alignment benchmark \cite{gold_deen-koehn-2005-europarl, vilar-etal-2006-aer} consisting of 508 manually annotated Europarl sentence pairs with sure and possible alignment links. Table~\ref{tab:aer} summarizes Alignment Error Rate (AER), precision, recall, total alignment links, and F-measure across five symmetrization strategies.

\begin{table}[!hbp]
\centering
\footnotesize
\setlength{\tabcolsep}{4pt}
\begin{tabular}{lcccc}
\toprule
\textbf{Symm.} & \textbf{AER} $\downarrow$ & \textbf{P} & \textbf{R} & \textbf{F} \\
\midrule
GD    & 37.4\% & 66.7\% & 58.8\% & 0.625 \\
GDFA  & 37.7\% & 65.2\% & 59.6\% & 0.623 \\
Inter.& 39.6\% & \textbf{82.2\%} & 47.6\% & 0.603 \\
GDF   & 40.0\% & 58.4\% & 61.7\% & 0.600 \\
Union & 40.8\% & 56.3\% & 62.6\% & 0.593 \\
\bottomrule
\end{tabular}
\caption{FastAlign symmetrization heuristics evaluated on the RWTH German--English gold alignment \cite{gold_deen-koehn-2005-europarl, vilar-etal-2006-aer} benchmark. GD: grow-diag; GDFA: grow-diag-final-and; GDF: grow-diag-final; Inter.: intersection. P, R, and F denote precision, recall, and F-measure. Bold denotes the highest precision. Intersection symmetrization is used in the prosody pipeline due to its substantially higher precision.}
\label{tab:aer}
\vspace{-3mm}
\end{table}

Although grow-diag achieves the lowest overall AER, intersection produces substantially higher precision than all other heuristics. For the downstream prosody pipeline, alignment precision is more important than maximizing alignment coverage. Each alignment link determines which source and target word-level prosodic segments are paired during cross-lingual contour comparison. Incorrect alignment links therefore introduce mismatched prosodic frames that directly corrupt the correlation signal. In contrast, missing alignments primarily reduce the number of usable word clusters without introducing additional noise.

We therefore adopt intersection symmetrization despite its moderately higher AER. The substantially improved precision (82.2\%) provides more reliable semantic correspondence between aligned prosodic segments, which is preferable for downstream fine-grained prosodic analysis.

\begin{table*}[!t]
\centering
\small
\setlength{\tabcolsep}{2.5pt}
\begin{tabular}{llcccc}
\toprule
\textbf{Feature} & \textbf{Comparison} & \textbf{Mean diff.} & \textbf{95\% CI} & \textbf{Cliff's $\delta$} & \textbf{$p$ (Holm)} \\
\midrule
Pitch  & DE--EN vs.\ EN--ES & $-0.0114$ & [$-0.020$, $-0.003$] & $-0.020$ & $3\times10^{-3}$ \\
Pitch  & DE--EN vs.\ EN--FR & $-0.0165$ & [$-0.026$, $-0.007$] & $-0.029$ & $2\times10^{-5}$ \\
Pitch  & EN--ES vs.\ EN--FR & $-0.0051$ & [$-0.014$, $+0.004$] & $-0.009$ & $0.16$ \\
\midrule
Energy & DE--EN vs.\ EN--ES & $+0.0092$ & [$+0.004$, $+0.015$] & $+0.031$ & $3\times10^{-6}$ \\
Energy & DE--EN vs.\ EN--FR & $+0.0074$ & [$+0.002$, $+0.013$] & $+0.023$ & $7\times10^{-4}$ \\
Energy & EN--ES vs.\ EN--FR & $-0.0019$ & [$-0.008$, $+0.003$] & $-0.008$ & $0.18$ \\
\bottomrule
\end{tabular}
\caption{Pairwise differences in mean prosodic correlation between language pairs. Confidence intervals are obtained using episode-clustered bootstrap resampling. $p$-values are Holm-corrected.}
\label{tab:pairwise_language}
\vspace{-3mm}
\end{table*}

\section{Additional Correlation Analysis}
\label{app:additional_prosodic_correlation}

We provide additional statistical analyses of the prosodic correlations reported in Section~\ref{sec:prosody_results}. We first examine whether differences in mean correlation across language pairs are statistically and practically meaningful. We then directly compare pitch and energy correlations within the same utterances.

Table~\ref{tab:pairwise_language} reports pairwise differences in mean correlation between language pairs. Several pairwise differences are statistically significant; however, the corresponding effect sizes are negligible, with $|\text{Cliff's }\delta| \leq 0.031$. In particular, DE--EN exhibits slightly lower pitch correlation than EN--ES and EN--FR, while its energy correlation is slightly higher. The small effect sizes indicate that these differences are of limited practical magnitude and that the overall degree of prosodic correspondence is similar across the three analyzed language pairs.

Table~\ref{tab:pitch_energy_comparison} directly compares pitch and energy correlations within the same utterances. Pitch correlation is higher than energy correlation on average for all three language pairs, with mean paired differences ranging from 0.046 to 0.070. Pitch correlation exceeds energy correlation in 57--60\% of individual utterances. The episode-clustered bootstrap confidence intervals for the mean paired differences exclude zero in every language pair, and paired Wilcoxon signed-rank tests show significant differences in all three cases.

\begin{table}[!tbp]
\centering
\footnotesize
\setlength{\tabcolsep}{4pt}
\begin{tabular}{lccccc}
\toprule
\textbf{Pair} & \textbf{Pitch} & \textbf{Energy} & \textbf{Diff.} & \textbf{95\% CI} & \textbf{P$>$E} \\
\midrule
DE--EN & 0.229 & 0.183 & $+0.046$ & [0.040, 0.052] & 57\% \\
EN--ES & 0.241 & 0.174 & $+0.067$ & [0.061, 0.074] & 60\% \\
EN--FR & 0.246 & 0.176 & $+0.070$ & [0.063, 0.077] & 60\% \\
\bottomrule
\end{tabular}
\caption{Paired comparison of pitch and energy correlations within the same utterances. Confidence intervals are obtained using episode-clustered bootstrap resampling of the mean paired difference. The P$>$E column reports the percentage of utterances for which pitch correlation exceeds energy correlation. Paired Wilcoxon signed-rank tests give $p < 10^{-80}$ for all three pairs.}
\label{tab:pitch_energy_comparison}
\vspace{-5mm}
\end{table}

Together, these analyses provide additional statistical support for the main prosodic correlation results. Although some differences between language pairs are statistically detectable due to the large sample size, their effect sizes are negligible. In contrast, the difference between pitch and energy correspondence is larger and consistent across all three language pairs, supporting the observation that pitch exhibits stronger cross-lingual correspondence than energy in the analyzed data.

\section{Additional POS Class Ablation Results}
\label{app:lexical_results}

The number of English words assigned to each POS class in the utterances is reported in Table~\ref{tab:pos_counts}.

\begin{table}[!hbp]
\centering
\footnotesize
\setlength{\tabcolsep}{5pt}
\begin{tabular}{lrrr}
\toprule
\textbf{POS group} & \textbf{DE--EN} & \textbf{EN--ES} & \textbf{EN--FR} \\
\midrule
Nouns     & 39,953  & 42,961  & 39,782  \\
Verbs     & 55,572  & 58,861  & 56,067  \\
Modifiers & 25,575  & 25,774  & 24,009  \\
Function  & 100,125 & 106,768 & 100,662 \\
Other     & 4,420   & 4,696   & 4,224   \\
\midrule
Total     & 225,645 & 239,060 & 224,744 \\
\bottomrule
\end{tabular}
\caption{English word counts per POS group in the scored utterances.}
\label{tab:pos_counts}
\vspace{-4mm}
\end{table}

Table~\ref{tab:appendix_full_lexical} provides the full leave-one-group-out POS-group ablation statistics, including mean, median, standard deviation, and the number of utterances containing each group. Pitch effects are more heterogeneous across utterances, whereas energy effects are more consistent, particularly for nouns and function words.

\begin{table*}[!tp]
\centering
\scriptsize
\begin{tabular}{llrrrrrrrr}
\toprule
Feature & Group & Pair & $n_{\text{total}}$ & $n_{\text{group}}$ & $r_{\text{full}}$ & $r_{\text{without}}$ & $\Delta r_{\text{mean}}$ & $\Delta r_{\text{median}}$ & $\Delta r_{\text{std}}$ \\
\midrule
\multirow{12}{*}{Pitch}
& Nouns     & DE--EN & 16,335 & 16,170 & 0.2295 & 0.2063 & +0.0228 & 0.0000 & 0.2099 \\
& Nouns     & EN--ES & 16,939 & 16,753 & 0.2409 & 0.2206 & +0.0208 & 0.0000 & 0.2273 \\
& Nouns     & EN--FR & 15,957 & 15,815 & 0.2460 & 0.2259 & +0.0203 & 0.0000 & 0.2132 \\
\addlinespace
& Verbs     & DE--EN & 16,335 & 16,236 & 0.2295 & 0.2262 & +0.0034 & 0.0000 & 0.1983 \\
& Verbs     & EN--ES & 16,939 & 16,756 & 0.2409 & 0.2261 & +0.0156 & 0.0000 & 0.2187 \\
& Verbs     & EN--FR & 15,957 & 15,915 & 0.2460 & 0.2416 & +0.0045 & 0.0000 & 0.1751 \\
\addlinespace
& Modifiers & DE--EN & 16,335 & 16,269 & 0.2295 & 0.2212 & +0.0083 & 0.0000 & 0.1499 \\
& Modifiers & EN--ES & 16,939 & 16,866 & 0.2409 & 0.2355 & +0.0057 & 0.0000 & 0.1528 \\
& Modifiers & EN--FR & 15,957 & 15,911 & 0.2460 & 0.2409 & +0.0055 & 0.0000 & 0.1438 \\
\addlinespace
& Function  & DE--EN & 16,335 & 16,258 & 0.2295 & 0.2368 & -0.0065 & 0.0000 & 0.2451 \\
& Function  & EN--ES & 16,939 & 16,795 & 0.2409 & 0.2305 & +0.0108 & 0.0000 & 0.2394 \\
& Function  & EN--FR & 15,957 & 15,827 & 0.2460 & 0.2327 & +0.0132 & 0.0000 & 0.2686 \\
\midrule
\multirow{12}{*}{Energy}
& Nouns     & DE--EN & 16,335 & 16,197 & 0.1831 & 0.1468 & +0.0355 & +0.0063 & 0.1303 \\
& Nouns     & EN--ES & 16,939 & 16,801 & 0.1739 & 0.1339 & +0.0396 & +0.0076 & 0.1382 \\
& Nouns     & EN--FR & 15,957 & 15,842 & 0.1758 & 0.1352 & +0.0402 & +0.0126 & 0.1297 \\
\addlinespace
& Verbs     & DE--EN & 16,335 & 16,279 & 0.1831 & 0.1875 & -0.0041 & 0.0000 & 0.1271 \\
& Verbs     & EN--ES & 16,939 & 16,853 & 0.1739 & 0.1805 & -0.0064 & 0.0000 & 0.1304 \\
& Verbs     & EN--FR & 15,957 & 15,908 & 0.1758 & 0.1886 & -0.0128 & -0.0005 & 0.1391 \\
\addlinespace
& Modifiers & DE--EN & 16,335 & 16,290 & 0.1831 & 0.1803 & +0.0027 & 0.0000 & 0.0916 \\
& Modifiers & EN--ES & 16,939 & 16,897 & 0.1739 & 0.1677 & +0.0062 & 0.0000 & 0.0921 \\
& Modifiers & EN--FR & 15,957 & 15,934 & 0.1758 & 0.1706 & +0.0052 & 0.0000 & 0.0891 \\
\addlinespace
& Function  & DE--EN & 16,335 & 16,297 & 0.1831 & 0.2156 & -0.0324 & -0.0211 & 0.1711 \\
& Function  & EN--ES & 16,939 & 16,886 & 0.1739 & 0.1943 & -0.0203 & -0.0093 & 0.1750 \\
& Function  & EN--FR & 15,957 & 15,928 & 0.1758 & 0.1961 & -0.0204 & -0.0084 & 0.1634 \\
\bottomrule
\end{tabular}
\caption{Full leave-one-group-out POS-group ablation statistics across language pairs and prosodic dimensions. $n_{\text{group}}$ denotes the number of utterances containing at least one aligned word from the corresponding POS group. Positive $\Delta r$ values indicate that removing the group decreases overall utterance-level correlation, suggesting that the group has a positive effect on cross-lingual prosodic correspondence.}
\label{tab:appendix_full_lexical}
\end{table*}

Episode-clustered bootstrap confidence intervals in Table~\ref{tab:pos_bootstrap_ci} show that noun-associated regions have reliably positive ablation effects across all language pairs for both pitch and energy, while function-word effects on energy are consistently negative.

\begin{table*}[!tp]
\centering
\small
\setlength{\tabcolsep}{6pt}
\begin{tabular}{llccc}
\toprule
\textbf{Feature} & \textbf{Group} & \textbf{DE--EN 95\% CI} & \textbf{EN--ES 95\% CI} & \textbf{EN--FR 95\% CI} \\
\midrule
Pitch & Nouns     & [+0.0193, +0.0266] & [+0.0180, +0.0253] & [+0.0166, +0.0235] \\
Pitch & Verbs     & [-0.0003, +0.0049] & [+0.0119, +0.0181] & [+0.0023, +0.0074] \\
Pitch & Modifiers & [+0.0062, +0.0105] & [+0.0034, +0.0080] & [+0.0031, +0.0079] \\
Pitch & Function  & [-0.0098, -0.0018] & [+0.0076, +0.0149] & [+0.0100, +0.0185] \\
\midrule
Energy & Nouns     & [+0.0331, +0.0370] & [+0.0378, +0.0420] & [+0.0377, +0.0421] \\
Energy & Verbs     & [-0.0004, +0.0031] & [-0.0085, -0.0042] & [-0.0080, -0.0044] \\
Energy & Modifiers & [+0.0012, +0.0042] & [+0.0048, +0.0076] & [+0.0038, +0.0065] \\
Energy & Function  & [-0.0445, -0.0384] & [-0.0222, -0.0169] & [-0.0332, -0.0273] \\
\bottomrule
\end{tabular}
\caption{Episode-clustered bootstrap 95\% confidence intervals for the leave-one-group-out POS-group ablation effects. An interval excluding zero indicates a statistically detectable ablation effect.}
\label{tab:pos_bootstrap_ci}
\end{table*}

Table~\ref{tab:noun_function} further reports direct paired noun--function comparisons, showing a robust noun--function contrast for all energy analyses and for DE--EN and EN--ES pitch, while the EN--FR pitch contrast is less robust under episode-level clustering.

\begin{table*}[!tp]
\centering
\small
\setlength{\tabcolsep}{6pt}
\begin{tabular}{llcccc}
\toprule
\textbf{Feature} & \textbf{Pair} & \textbf{Noun--Function} & \textbf{95\% CI} & \textbf{$p$} & \textbf{Result} \\
\midrule
Pitch & DE--EN & +0.0285 & [+0.022, +0.035] & $2\times10^{-26}$ & Significant \\
Pitch & EN--ES & +0.0104 & [+0.005, +0.016] & $2\times10^{-4}$ & Significant \\
Pitch & EN--FR & +0.0056 & [-0.001, +0.012] & $2\times10^{-3}$ & Marginal \\
\midrule
Energy & DE--EN & +0.0771 & [+0.073, +0.081] & $<10^{-300}$ & Significant \\
Energy & EN--ES & +0.0598 & [+0.056, +0.064] & $3\times10^{-226}$ & Significant \\
Energy & EN--FR & +0.0706 & [+0.066, +0.075] & $2\times10^{-261}$ & Significant \\
\bottomrule
\end{tabular}
\caption{Paired comparison of noun and function-word POS-group ablation effects. Confidence intervals are obtained using episode-clustered bootstrap resampling; $p$-values are from paired Wilcoxon signed-rank tests.}
\label{tab:noun_function}
\end{table*}

\section{Temporal Resampling and F0 Interpolation Robustness}
\label{app:temporal_robustness}

We conduct additional robustness analyses to examine whether the reported prosodic correlations are sensitive to temporal resampling and F0 interpolation choices. First, we evaluate different values of frames per word used for contour resampling. Table~\ref{tab:fpw_sensitivity} shows that the mean per-utterance zero-lag Spearman correlations remain nearly unchanged across values from 10 to 40 frames per word. The maximum spread across settings is 0.0008, indicating that the choice of 20 frames per word does not affect the reported conclusions. The absolute correlation values are computed on a random subset of 3,000 utterance pairs per language pair and may differ slightly from the full-data results; the stability across settings is the relevant robustness measure.

\begin{table}[!tbp]
\centering
\footnotesize
\setlength{\tabcolsep}{3pt}
\begin{tabular}{lcccccc}
\toprule
\textbf{Pair} & \textbf{10} & \textbf{15} & \textbf{20} & \textbf{30} & \textbf{40} & \textbf{Spread} \\
\midrule
EN--ES & 0.2496 & 0.2498 & 0.2495 & 0.2496 & 0.2498 & 0.0002 \\
DE--EN & 0.2310 & 0.2310 & 0.2315 & 0.2318 & 0.2318 & 0.0008 \\
EN--FR & 0.2448 & 0.2444 & 0.2451 & 0.2450 & 0.2450 & 0.0007 \\
\bottomrule
\end{tabular}
\caption{Robustness of mean per-utterance Spearman correlation to different frames-per-word values (columns) using interpolated F0 contours. Each value is computed on a random subset of 3,000 utterance pairs per language pair.}
\label{tab:fpw_sensitivity}
\end{table}

We additionally examine whether interpolation over unvoiced regions affects the measured pitch correspondence. Since F0 is only defined during voiced speech, directly comparing raw F0 contours can introduce additional mismatch caused by language-specific differences in phonetic realization and voiced/unvoiced distributions. Table~\ref{tab:f0_interpolation} compares interpolated and raw F0 contours for the EN--ES language pair across different frames-per-word settings. Raw contours yield lower correlations, while both variants remain stable across temporal resolutions. Different languages naturally exhibit different phoneme sequences and consequently different distributions of voiced and unvoiced regions, so without interpolation the measured correspondence would be disproportionately influenced by these differences even when the underlying intonational contour is similar.

\begin{table}[!tbp]
\centering
\footnotesize
\setlength{\tabcolsep}{4pt}
\begin{tabular}{lccccc}
\toprule
\textbf{Variant} & \textbf{10} & \textbf{15} & \textbf{20} & \textbf{30} & \textbf{40} \\
\midrule
Interpolated & 0.2317 & 0.2316 & 0.2315 & 0.2315 & 0.2318 \\
Raw          & 0.1675 & 0.1668 & 0.1671 & 0.1670 & 0.1666 \\
\bottomrule
\end{tabular}
\caption{Comparison of interpolated and raw F0 correlations for EN--ES across frames-per-word values (columns). The maximum spread across settings is 0.0003 for interpolated and 0.0009 for raw contours. Results are evaluated on a subset of 3,000 utterance pairs.}
\label{tab:f0_interpolation}
\end{table}

\section{Gender Distribution}

Speaker gender distribution was estimated using an automatic speech-based gender classifier \cite{gendertool} applied to each utterance; the resulting proportions should therefore be interpreted as approximate. Across utterances, English speech is 57.7\% male and 42.3\% female, French 54.8\% / 45.2\%, German 54.2\% / 45.8\%, and Spanish 57.7\% / 42.3\%.

\end{document}